\documentclass{aiaa-tc}

 \usepackage{varioref}
 \usepackage{wrapfig}
 \usepackage{threeparttable}
 \usepackage{dcolumn}
  \newcolumntype{d}{D{.}{.}{-1}}
 \usepackage{nomencl}
 \usepackage{subfigure}
 \usepackage{subfigmat}
 \usepackage{fancyvrb}
  \fvset{fontsize=\footnotesize,xleftmargin=2em}
 \usepackage{lettrine}
 \usepackage{hyperref}
 \usepackage{setspace}
 \usepackage{amssymb}
 \usepackage{amsmath}
 \makenomenclature
 \title{A Quasi-one-dimensional Analytic Model of Rotating Detonation Combustors}

 \author{
  Rei Kawashima\thanks{Assistant Professor, Department of Aeronautics and Astronautics, 7-3-1 Hongo, Bunkyo-ku. Member AIAA.}\\
  {\normalsize\itshape
   The University of Tokyo, Tokyo 113-8656, Japan}\\
  Ikkoh Funaki\thanks{Associate Professor, Institute of Space and Astronautical Science, 3-1-1 Yoshinodai, Sagamihara. Member AIAA.}\\
  {\normalsize\itshape
   Japan Aerospace Exploration Agency, Kanagawa 252-5210, Japan}\\
  Jumpei Fujii\thanks{Graduate Student, School of Science for Open and Environmental Systems, 3-14-1 Hiyoshi, Kohoku-ku, Yokohama. Student Member AIAA.}
   and Akiko Matsuo\thanks{Professor, School of Science for Open and Environmental Systems, 3-14-1 Hiyoshi, Kohoku-ku, Yokohama. Senior Member AIAA.}\\
  {\normalsize\itshape
   Keio University, Kanagawa 223-8522, Japan}\\
  and\\
  Kazuki Ishihara\thanks{Graduate Student, Department of Aerospace Engineering, Furo-cho, Chikusa-ku, Nagoya. Student Member AIAA.}
  and Jiro Kasahara\thanks{Professor, Department of Aerospace Engineering, Furo-cho, Chikusa-ku, Nagoya. Senior Member AIAA.}\\
  {\normalsize\itshape
   Nagoya University, Aichi 464-8603, Japan}\\
 }

 \AIAApapernumber{200?-????}
 \AIAAconference{Conference Name, Date, and Location}
 \AIAAcopyright{\AIAAcopyrightD{200?}}

\begin{document}

\maketitle

\begin{abstract}
	A quasi-one-dimensional analytic model is proposed for the internal fluid of rotating detonation combustors (RDCs).
	This model uses the shock-tube model that constrains the flow to have only a longitudinal component, while admitting the propagation of the detonation wave in the azimuthal direction.
	The proposed model is able to compute the thruster performance and two-dimensional distributions of gas properties.
	The calculation process of the model excludes iterative calculation or space discretization.
	The case calculations of the hydrogen-air RDC and the ethylene-oxygen RDC are conducted, and the results calculated by the analytic model are compared with those simulated by computational fluid dynamics (CFD).
	Good agreement has been observed between the results obtained with the proposed model and CFD, in terms of both of the qualitative and quantitative comparisons.
	The proposed model is simple and fast, and also maintains the fundamental characteristics of RDCs.
\end{abstract}

\printnomenclature

\nomenclature[1a]{$A$}{area}
\nomenclature[1a]{$a$}{acoustic velocity}
\nomenclature[1d]{$D$}{detonation speed}
\nomenclature[1f]{$f$}{wall pressure decay function}
\nomenclature[1h]{$h$}{detonation wave height}
\nomenclature[1l]{$L_{\rm c}$}{axial annulus length}
\nomenclature[1l]{$L_{\rm \theta}$}{azimuthal annulus length}
\nomenclature[1w]{$w_{\rm c}$}{annulus width}
\nomenclature[1n]{$N$}{detonation wave number}
\nomenclature[1p]{$P$}{pressure}
\nomenclature[1t]{$T$}{temperature}
\nomenclature[1k]{$k$}{coefficient in pressure decay function}
\nomenclature[1r]{$R$}{gas constant}
\nomenclature[1rpg]{$r_{\rm PG}$}{pressure gain ratio}
\nomenclature[1m]{$\dot{m}$}{mass flow rate}
\nomenclature[1f]{$F$}{thrust force}
\nomenclature[1t]{$t$}{time}
\nomenclature[1u]{$u$}{velocity}
\nomenclature[1isp]{$I_{\rm sp}$}{specific impulse}
\nomenclature[2c]{$\gamma$}{specific heat ratio}
\nomenclature[2l]{$\lambda$}{detonation cell width}
\nomenclature[ba]{a}{atomospheric}
\nomenclature[ba]{ave}{average}
\nomenclature[bb]{b}{burned gas}
\nomenclature[bc]{c}{choked}
\nomenclature[bcj]{CJ}{Chapman-Jouguet condition}
\nomenclature[bi]{i}{injector}
\nomenclature[bp]{p}{plenum}
\nomenclature[bu]{u}{unburned gas}
\nomenclature[bw]{w}{wall}
\nomenclature[beff]{eff}{effective}
\nomenclature[b1]{0}{unburned gas ahead of detonation}
\nomenclature[b2]{1}{low-pressure room burned gas}
\nomenclature[b3]{2}{high-pressure room burned gas}

\section{Introduction}

	The rotating detonation combustor (RDC) is a continuous detonation combustor.
	The RDC has several advantages, such as pressure gain combustion and compactness, compared with conventional constant-pressure combustors.
	Owing to these advantages, the applications of the RDC to the rocket engine, jet engine, and gas turbine are envisaged as discussed in Ref. [\citen{Lu:2014aa}].
	In order to conduct an engine system analysis using the RDC, the combustor must be modeled.
	Because the internal fluid of the RDC is nonuniform in the azimuthal direction, the modeling must consider the azimuthal distribution.
	Further, to predict the heat fluxes from the combustion gas to the walls, the model should also consider the axial distribution.
	Therefore, for accurate analysis of the systems using the RDC, it is necessary to construct an azimuthal-axial two-dimensional model for the internal fluid of the RDC annulus. 

	The two-dimensional distributions of the RDC internal fluid are discussed using several high-fidelity numerical simulations of computational fluid dynamics (CFD).
	Schwer and Kailasanath have conducted two-dimensional simulations of the hydrogen-air RDCs, and discussed the characteristics and structure of the internal fluid [\citen{Schwer:2011aa}]. 
	Fujii et al. have conducted a two-dimensional simulation of an ethylene-oxygen RDC to investigate the effect of the injector configuration on the internal fluid [\citen{Fujii:2017aa}].
	These numerical simulations yield a lot of knowledge about the physics of the internal fluid and the thruster performance of the RDC.
	However, these numerical simulations cannot be integrated into the engine system analysis as one component, because a long calculation time is required for the analysis.
	A simplified model is required to enable a fast analysis of the RDC, while maintaining the fundamental physics of the internal fluid observed in the CFD analyses.

	There have been several studies on theoretical models for the RDC internal fluid.
	Braun et al. constructed a theoretical model of the RDC internal fluid by using the empirical equation on the detonation wave height [\citen{Bykovskii:2006aa}], and conducted a parametric study for the RDC performance [\citen{BraunAST2013}].
	Kaemming et al. built a model using an empirical equation of axial flow velocity derived by CFD [\citen{Kaemming:2016aa}].
	These models have been validated by comparing the results obtained by the model and CFD.
	These theoretical models are beneficial in the sense that they can quickly analyze the performance of the RDC.
	However, these models do not consider the axial distribution of the internal fluid.
	There has been no analytic model that considers both the azimuthal and axial distributions.
	As an intermediate approach between CFD and theoretical model, a two-dimensional analysis using the method of characteristics has been conducted by Fievisohn et al. [\citen{Fievisohn:2016ab}].
	This model basically consists of the method of characteristics, the Chapman-Jouguet detonation solver, and the fundamental equations of oblique shock waves.
	A parametric study was conducted by changing the combustor shape and equivalence ratio, and it was reported that the results obtained by the model agree well with the results of CFD.
	This model is advantageous in the sense that it can analyze two-dimensional distributions.
	However, it requires several cumbersome processes such as spatial discretization, compared with theoretical models.
	
	In this paper, an analytic model for the internal fluid of the RDC is considered,
	which is characterized by the following properties:
	1) a theoretical model that does not require any iterative calculation or space discretization,
   and 2) that has the capability of analyzing azimuthal-axial 2D distributions of the internal fluid.
	It is difficult to consider a fully 2D model for the internal fluid.
	Thus, a quasi-one-dimensional (quasi-1D) model is considered that admits the propagation of the detonation wave in the azimuthal direction, while the flow is constrained to have only an axial component.
	
\section{A Quasi-one-dimensional Model of RDC}
\label{sec:model}

\subsection{Model Overview}
\begin{figure}[t]
		\begin{center}
			\includegraphics[width=63mm]{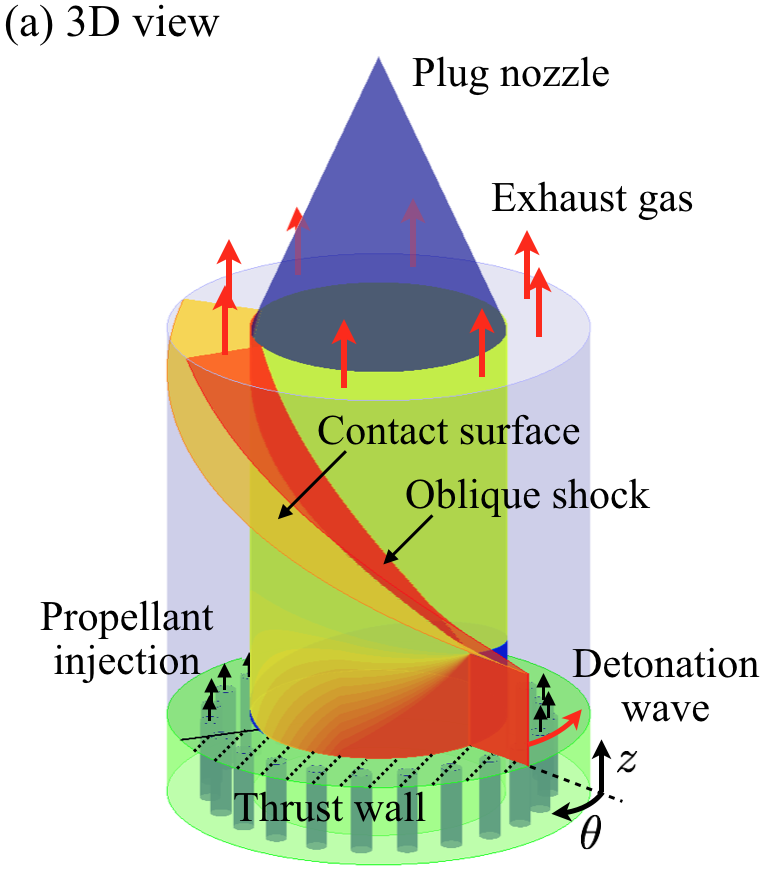}
		\end{center}
		\begin{center}
			\includegraphics[width=80mm]{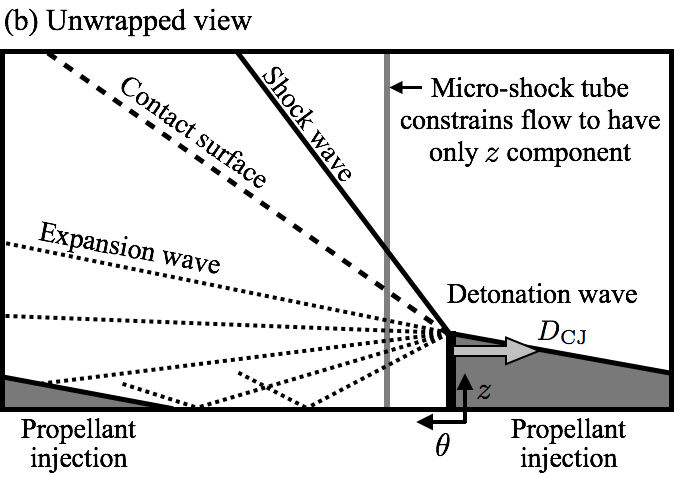}
		\end{center}
		\caption{Conceptual figures of the RDC assumed in the quasi-1D model. (a) Three-dimensional view, and (b) unwrapped view. 
		}
		\label{fig:concept}
	\end{figure}

	The conceptual figure of an annular-type RDC is shown in Fig. \ref{fig:concept}(a).
	The model is simplified to handle the azimuthal-axial 2D distributions in the unwrapped domain, as shown in Fig. \ref{fig:concept}(b) .
	The wave structure consisting of the oblique shock, contact discontinuity, and expansion wave is observed in the internal flow of the RDC.
	If the detonation wave is assumed to be propagating in the azimuthal direction with a constant speed, the azimuthal position can be interpreted as the time axis after the passing of the detonation wave.
	Hence, it is assumed that the wave structure of the RDC internal flow resembles that of a typical shock tube.
	On the basis of this assumption, the concept of the quasi-1D model is considered.
	This model assumes the micro-shock tubes that constrain the flow to have only an axial component.
	The wave structure in the RDC is approximated by the wave propagation processes in a shock tube.
	Although the azimuthal flow is not induced in this model, the detonation combustion wave is propagating in the azimuthal direction.
	On the basis of this concept, the quasi-1D model is constructed by using the theoretical formulas of one-dimensional flows.
	The thrust force is calculated by the pressure history at the thrust wall.
	The thrust estimation method based on the wall pressure history has been used in the analytic models for pulse detonation engines [\citen{Wintenberger:2003aa,Zitoun:1999aa}].
	As well as the standard CFD analyses [\citen{Schwer:2011aa,Fujii:2017aa}], the input parameters used in this model are set as follows: fuel species, oxidizer species, azimuthal annulus length $L_{\theta}$, axial annulus length $L_{\rm c}$, channel width $w_{\rm c}$, plenum pressure $P_{\rm p}$, plenum temperature $T_{\rm p}$, injector-wall area ratio $A_{\rm i}/A_{\rm w}$, and background pressure $P_{\rm b}$.

\subsection{Injector Model}
	\label{sec:injector}
\begin{figure}[t]
		\begin{center}
			\includegraphics[width=70mm]{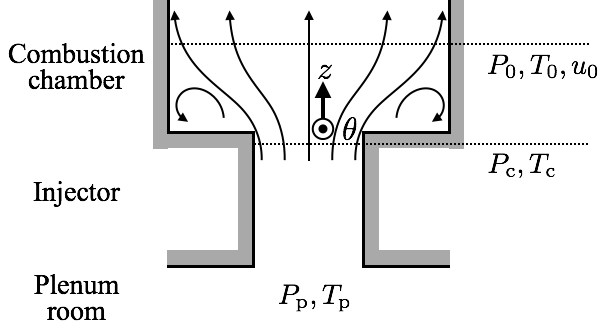}
		\end{center}
		\caption{Schematic of plenum room, injector, and annulus combustion chamber.}
		\label{fig:injector}
	\end{figure}

	First, the injector model is considered to determine the characteristics of the unburned propellants ahead of the detonation.
	The schematic of a plenum room, injector, and annulus combustion chamber is shown in Fig. \ref{fig:injector}.
	It has been assumed that the fuel and oxidizer are perfectly premixed before injection.
	In what follows, the premixture of the fuel and oxidizer is called the propellant, which has a pressure and temperature of $P_{\rm p}$ and $T_{\rm p}$ in the plenum room.
	It is further assumed that the flow is always choked at the injector exit surface when the propellant is injected.
	The pressure and temperature at the injector exit can be expressed as follows:	
	\begin{equation}
		P_{\rm c}=\left(\frac{2}{\gamma_{\rm u}+1}\right)^{\frac{\gamma_{\rm u}}{\gamma_{\rm u}-1}}P_{\rm p},\hspace{20pt}T_{\rm c}=\left(\frac{2}{\gamma_{\rm u}+1}\right)T_{\rm p}.
	\end{equation}
	Here, the pressure of the choked flow $P_{\rm c}$ is also called the ``critical pressure.''
	
	The pressure of the unburned propellant injected into the annulus further decreases from $P_{\rm c}$ owing to the injector-wall area ratio $A_{\rm i}/A_{\rm w}$.
	The cross-sectional area of the flow path rapidly increases at the exit of the injectors.
	In this case, the pressure after the area expansion is given as a function of $A_{\rm i}/A_{\rm w}$ and Mach number $M_0$, as follows:
	\begin{equation}
		P_0=\frac{A_{\rm i}}{A_{\rm w}}\frac{1}{M_0}
		\left[\frac{\gamma_{\rm u}+1}{2+\left(\gamma_{\rm u}-1\right)M_0^2}\right]^{\frac{1}{2}}P_{\rm c}.
		\label{eq:P0}
	\end{equation}
	In fact, it is difficult to determine the Mach number $M_0$ using only the upstream information.
	Here, $M_0$ is artificially selected as 0.65 to match $P_0$ with that simulated in CFD.
	By using $M_0$, $T_0$ and $u_0$ can also be calculated as follows:
	\begin{equation}
		T_{0}=\frac{\gamma_{\rm u}+1}{2+\left(\gamma_{\rm u}-1\right)M_0^2}T_{\rm c},
	\end{equation}
	\begin{equation}
		u_{0}=M_0\left[\frac{\gamma_{\rm u}+1}{2+\left(\gamma_{\rm u}-1\right)M_0^2}\right]^{\frac{1}{2}}a_{\rm c}.
	\end{equation}

\subsection{Detonation Combustion Model}
	The chemical reaction calculation in the detonation combustion wave is conducted by using the AISTJAN [\citen{Tanaka2006}].
	The AISTJAN is a detonation characteristics calculator developed by the National Institute of Advanced Industrial Sciences and Technology.
	For the pressure and temperature before the combustion, $P_0$ and $T_0$ are used.
	
	In the quasi-1D model, a special treatment is required in the combustion calculation for consistency with the model concept.
	Because the azimuthal flow is not induced after the detonation wave, the combustion process assumed in this model should be constant-volume (CV) combustion, rather than detonation combustion.
	The properties of the burned gas, such as $P_2$, $T_2$, and $a_2$, are calculated by assuming CV combustion in the AISTJAN.
	Nevertheless, the detonation wave speed is required in the model.
	The Chapman-Jouguet (CJ) detonation calculation is implemented for deriving the detonation wave speed $D_{\rm CJ}$.
	It is noted that the detonation speeds measured in experiments are usually slower than $D_{\rm CJ}$ [\citen{Kudo:2011aa,Kato:2016aa}].
	However, the detonation wave speeds simulated in the CFDs are quite close to $D_{\rm CJ}$, especially in the cases of premixed propellant.
	Because the quasi-1D model aims at a theoretical model that simplifies the CFD analyses, $D_{\rm CJ}$ is used for the detonation propagation speed.

\subsection{Wall Pressure and Detonation Wave Height Model}
	\label{sec:wall}
	
\begin{figure}[t]
		\begin{center}
			\includegraphics[width=85mm]{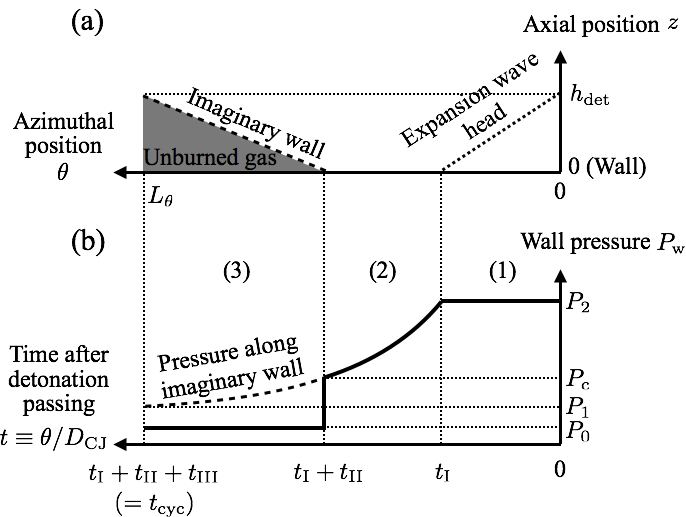}
		\end{center}
		\caption{(a) Wave structure in front of the annulus wall.
		(b) Time history of the annulus wall pressure.
		}
		\label{fig:wallmodel}
	\end{figure}

	The wall pressure distribution is considered within one cycle of rotating detonation.
	The concept of the wall pressure model is illustrated in Fig. \ref{fig:wallmodel}.
	The cycle period of rotating detonation is defined as $t_{\rm cyc}\equiv L_\theta/\left(N_{\rm det}D_{\rm CJ}\right)$.
	Here, $t_{\rm cyc}$ is divided into three sections as follows:
	\begin{equation}
		t_{\rm cyc}=t_{\rm I}+t_{\rm II}+t_{\rm III}.
		\label{eq:timecyc}
	\end{equation}
	$t_{\rm I}$ is the time for the expansion wave head to reach the annulus wall after the detonation wave passes. 
	$t_{\rm II}$ is the time for the wall pressure to decay to the critical pressure of the injector after $t=t_{\rm I}$.
	$t_{\rm III}$ is the time during which the propellant is injected into the annulus.
	The azimuthal distribution of wall pressure $P_{\rm w}$ in each section is considered to be related to the detonation wave height $h_{\rm det}$.
	In what follows, the azimuthal position is regarded as the time after the detonation wave passes, and $t=0$ is defined at the position of the detonation wave.
	\\
	\noindent(1) $0<t\leq t_{\rm I}$
	
	The speed of the rarefaction wave head propagating toward the wall is assumed to be the same as the acoustic velocity of the burned gas.
	Thus, $t_{\rm I}$ is calculated as follows:
	\begin{equation}
		t_{\rm I}=\frac{h_{\rm det}}{a_2}.
		\label{eq:time1}
	\end{equation}
	The wall pressure is maintained at the pressure of the high-pressure gas until the rarefaction wave head reaches the wall.
	\begin{equation}
		P_{\rm w}=P_2.
		\label{eq:pres1}
	\end{equation}

	\noindent(2) $t_{\rm I}<t\leq t_{\rm I}+t_{\rm II}$

	After the rarefaction wave head reaches the wall, rarefaction waves propagate to the wall and reflection waves travel to the exit.
	The pressure field in front of the wall becomes complicated owing to the interactions of these waves.
	The wall pressure decay model is discussed in the simplified theoretical pulse detonation engine (PDE) model by Endo et al. [\citen{Endo:2004aa}].
	According to Endo's simplified model, if one assumes that the rarefaction wave is self similar, analytic solutions of wall pressure decay can be derived for discrete specific heat ratios.
	Referring to Endo's PDE model, the relationship between the wall pressure and time is expressed as follows:
	\begin{equation}
		\frac{a_2\left(t-t_{\rm I}\right)}{h_{\rm det}}+1=f\left(\frac{P_2}{P_{\rm w}}\right).
		\label{eq:endo1}
	\end{equation}
	It is noted that $h_{\rm det}$ in Eq. (\ref{eq:endo1}) is replaced by the PDE tube length in Endo's PDE model.
	In the current RDC model, $h_{\rm det}$ is used because it is the propagation distance of the rarefaction wave head.
	The function $f$ is defined for discrete specific heat ratios and it is derived in Ref. [\citen{Endo:2004aa}] for several specific heat ratios.
	However, as mentioned in Ref. [\citen{Endo:2004aa}], the dependence of $f$ on the specific heat ratio is weak.
	Hence this analysis uses $f$ in the case of $\gamma=15/13$ as follows:
	\begin{align}
		f\left(x\right)=
		 &\frac{231}{2^{10}}x^{\frac{1}{15}}
		+\frac{63 }{2^{9 }}x^{\frac{3}{15}}
		+\frac{105}{2^{10}}x^{\frac{5}{15}}\nonumber \\
		+&\frac{25 }{2^{8 }}x^{\frac{7}{15}}
		+\frac{105}{2^{10}}x^{\frac{9}{15}}
		+\frac{63 }{2^{9 }}x^{\frac{11}{15}}
		+\frac{231}{2^{10}}x^{\frac{13}{15}}.
		\label{eq:endo2}
	\end{align}
	By using Eq. (\ref{eq:endo1}), $t_{\rm II}$ is expressed as follows:
	\begin{equation}
		t_{\rm II}=\frac{h_{\rm det}}{a_2}\left(f\left(\frac{P_2}{P_{\rm c}}\right)-1\right).
		\label{eq:time2}
	\end{equation}
	One can calculate $t_{\rm II}$ by combining Eqs. (\ref{eq:endo2}) and (\ref{eq:time2}).

	Endo et al. also give an alternative form of Eq. (\ref{eq:endo1}).
	$P_{\rm w}$ is expressed as an approximate function of time as follows:
	\begin{align}
		\frac{P_{\rm w}}{P_2}=&k_{\rm A}\exp\left[-k_{\rm B}\frac{a_2}{h_{\rm det}}\left(t-t_{\rm I}\right)\right]\nonumber \\
		+&\left(1-k_{\rm A}\right)\exp\left[-k_{\rm C}\frac{a_2}{h_{\rm det}}\left(t-t_{\rm I}\right)\right].
		\label{eq:pwall}
	\end{align}
	In the case of $\gamma=15/13$, the coefficients in this equation are approximated as follows:
	\begin{equation}
		k_{\rm A}=0.6066,\hspace{15pt}k_{\rm B}=2.991,\hspace{15pt}k_{\rm C}=0.5014.
		\label{eq:coef}
	\end{equation}
	Again, the dependence of Eq. (\ref{eq:pwall}) on the specific heat ratio is weak.
	Thus, Eqs. (\ref{eq:pwall}) and (\ref{eq:coef}) are used in the calculation of $P_{\rm w}$ for any combustion gases in this paper.
	\\
	\noindent(3) $t_{\rm I}+t_{\rm II}<t\leq t_{\rm I}+t_{\rm II}+t_{\rm III}=t_{\rm cyc}$
	
	It is assumed that the detonation wave height is achieved with the propellant within this period.
	$t_{\rm III}$ is then simply written as follows:
	\begin{equation}
		t_{\rm III}=\frac{h_{\rm det}}{u_0}.
		\label{eq:time3}
	\end{equation}
	During the propellant injection, the wall pressure is considered to be the same as the unburned propellant pressure. Therefore,
	\begin{equation}
		P_{\rm w}=P_{\rm 0}.
		\label{eq:pres3}
	\end{equation}
	
	It is noted that the boundary between the unburned and burned gases, denoted as the white dashed line in Fig. \ref{fig:wallmodel}, is called the deflagration region [\citen{Paxson:2014aa}].
	The flow field in this region is complicated because of its association with deflagration combustion.
	It is difficult to approximate the physics in this region by using only theoretical equations.
	Therefore, in this quasi-1D model, this boundary is treated as an imaginary wall, where the information convection between the unburned and burned gases is neglected.
	It is further assumed that the wall pressure decay model in Eq. (\ref{eq:pwall}) is extended to the imaginary wall.
	
	The detonation wave height is estimated by using $t_{\rm I}$, $t_{\rm II}$, and $t_{\rm III}$ deduced above.
	Substituting Eqs. (\ref{eq:time1}), (\ref{eq:time2}), and (\ref{eq:time3}) into Eq. (\ref{eq:timecyc}), $h_{\rm det}$ is described as follows:
	\begin{equation}
		h_{\rm det}=\frac{L_\theta}{N_{\rm det}D_{\rm CJ}}\left[\frac{1}{a_2}f\left(\frac{P_2}{P_{\rm c}}\right)+\frac{1}{u_0}\right]^{-1}.
		\label{eq:hdet}
	\end{equation}
	This estimation method for the detonation wave height does not require an empirical equation or parameter.
	The important model enabling this method is Endo's wall pressure decay model for PDEs.
	The validity of this method will be discussed in Sec. \ref{sec:case1}.
	
	As an alternative estimation method for the detonation wave height, Braun et al. uses an empirical equation related to the detonation cell width as follows [\citen{BraunAST2013}]:
	\begin{equation}
		h_{\rm det,min}=\left(12\pm5\right)\lambda_{\rm det}.
		\label{eq:detmin}
	\end{equation}
	Here, $h_{\rm det,min}$ is defined as the minimum value of $h_{\rm det}$ that can maintain the rotating detonation [\citen{Bykovskii:2006aa}].
	This method gives only the minimum value of the detonation wave height.
	Moreover, the uncertainty in Eq. (\ref{eq:detmin}) is approximately 40\%, which can cause erroneous predictions for the RDC performance.
	Thus, the theoretical method in Eq. (\ref{eq:hdet}) is used in this paper for prediction of the detonation height.
	
	For simplicity, the detonation wave number $N_{\rm det}$ is assumed to be unity in this paper.
	It would be difficult to determine the detonation wave number theoretically.
	It has been reported that the detonation wave number can be variable even in single operation condition of the RDC [\citen{Kato:2016aa}].
	To discuss the RDC operations in multiple wave mode, the empirical equation in Eq. (\ref{eq:detmin}) would be useful.
	For instance, if $h_{\rm det}$ as determined by Eq. (\ref{eq:hdet}) exceeds the range in Eq. (\ref{eq:detmin}), one may need to increase the detonation wave number $N_{\rm det}$.
	Wolanski has presented a method to predict the wave number based on this concept .
	A detailed consideration is needed for an accurate determination method of the detonation wave number, which is beyond the scope of this paper.

\subsection{Thruster Performance}
	The thruster performances are also theoretically derived.
	As described in Sec. \ref{sec:injector}, it is assumed that the propellant is injected only with the choked flow.
	The mass flow rate of the choked flow is as follows:
	\begin{equation}
		\dot{m} = P_{\rm p}A_{\rm i,eff}\sqrt{\frac{\gamma_{\rm u}}{RT_{\rm p}}}
		\left(\frac{\gamma_{\rm u}+1}{2}\right)^{-\frac{\gamma_{\rm u}+1}{2\left(\gamma_{\rm u}-1\right)}}.
	\end{equation}
	Here $A_{\rm i,eff}$ is the effective injector exit area where the choked flow is injected.
	As illustrated in Fig. \ref{fig:wallmodel}, the propellant is injected during $t_{\rm III}$, which corresponds to the azimuthal length of $D_{\rm CJ}t_{\rm III}$.
	Therefore, by using the injector-wall area ratio $A_{\rm i}/A_{\rm w}$, the effective injector exit area can be expressed as follows:
	\begin{equation}
		A_{\rm i,eff}=D_{\rm CJ}t_{\rm III}w_{\rm c}\frac{A_{\rm i}}{A_{\rm w}}.
	\end{equation}
	
	The thrust force is calculated by integrating the pressure distribution on the wall as follows:
	\begin{equation}
		F = \int_0^{A_{\rm w}}\left(P_{\rm w}-P_{\rm a}\right)dA
		= \int_0^{t_{\rm cyc}}P_{\rm w}D_{\rm CJ}w_{\rm c}dt-P_{\rm a}A_{\rm w}.
	\end{equation}
	As well as the division of the cycle period in Eq. (\ref{eq:timecyc}), the total thrust force is divided into the forces generated in the three sections, in addition to the ambient pressure contribution, as follows:
	\begin{equation}	
		F=F_{\rm I}+F_{\rm II}+F_{\rm III}-P_{\rm a}A_{\rm w},
	\end{equation}
	where $F_{\rm I}$, $F_{\rm II}$, and $F_{\rm III}$ correspond to the duration $t_{\rm I}$, $t_{\rm II}$, and $t_{\rm III}$, respectively.
	By using Eq. (\ref{eq:pres1}), $F_{\rm I}$ is calculated as follows:
	\begin{equation}
		F_{\rm I} = P_2D_{\rm CJ}t_{\rm I}w_{\rm c}.
	\end{equation}
	Integrating Eq. (\ref{eq:pwall}) with time, $F_{\rm II}$ is expressed as follows:
	\begin{align}
		F_{\rm II} =&\frac{P_2h_{\rm det}}{a_2}\left[\frac{k_{\rm A}}{k_{\rm B}}\left(1-\exp\left[-k_{\rm B}\frac{a_2}{h_{\rm det}}t_{\rm II}\right]\right)\right.\nonumber\\
		&\left.+\frac{1-k_{\rm A}}{k_{\rm C}}\left(1-\exp\left[-k_{\rm C}\frac{a_2}{h_{\rm det}}t_{\rm II}\right]\right)\right]D_{\rm CJ}w_{\rm c}.
	\end{align}
	By using Eq. (\ref{eq:pres3}), $F_{\rm III}$ is calculated as follows:
	\begin{equation}
		F_{\rm III} = P_0D_{\rm CJ}t_{\rm III}w_{\rm c}.
	\end{equation}
	Once the mass flow rate and thrust force are obtained, the specific impulse is calculated by $I_{\rm p}=F/\left(\dot{m}g\right)$.
	
	Another important factor may be how much pressure gain is achieved by using pressure gain combustion compared with constant-pressure combustion.
	The pressure gain ratio $r_{\rm PG}$ is used to evaluate the effectiveness of the rotating detonation combustion quantitatively.
	Here, $r_{\rm PG}$ is defined as the ratio of the average wall pressure to the unburned propellant pressure, which is calculated as follows:
	\begin{equation}
		r_{\rm PG}=\frac{P_{\rm w,ave}}{P_0}=\frac{F+P_{\rm a}A_{\rm w}}{P_0A_{\rm w}}.
	\end{equation}
	
\subsection{Shock-tube Model}
	To derive the 2D pressure distribution, the shock-tube model is used.
	The detailed equation sets for the wave propagations and spatial profiles can be found in a textbook [\citen{Zeldovich}], and hence they are not repeated here.
	In the shock-tube model, the pressures in the high-pressure room and low-pressure room are given as the initial condition.
	In the quasi-1D model, the pressure in the high-pressure room is given by $P_2$. 
	The pressure in the low-pressure room $P_1$ represents the pressure at the left-hand side end of the imaginary boundary in Fig. \ref{fig:wallmodel}. 
	Thus, by using Eq. (\ref{eq:pwall}), $P_1$ is calculated as follows:
	\begin{align}
		\frac{P_1}{P_2}=&k_{\rm A}\exp\left[-k_{\rm B}\frac{a_2}{h_{\rm det}}\left(t_{\rm cyc}-t_{\rm I}\right)\right]\nonumber \\
		+&\left(1-k_{\rm A}\right)\exp\left[-k_{\rm C}\frac{a_2}{h_{\rm det}}\left(t_{\rm cyc}-t_{\rm I}\right)\right].
	\end{align}
	
\subsection{Summary of the Model}
\begin{figure}[t]
		\begin{center}
			\includegraphics[width=85mm]{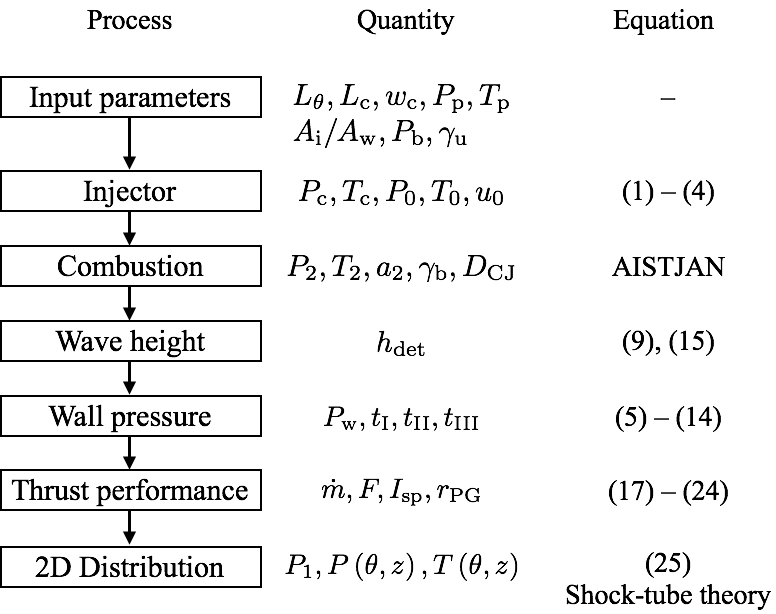}
		\end{center}
		\caption{Sequence of calculation processes.
		The quantities derived and equations used in each process are shown.
		}
		\label{fig:flow}
	\end{figure}

	One can compute the thruster performance and 2D distributions by following the procedures in Sec. \ref{sec:model}.
	The sequence of calculation processes and relevant quantities are summarized in Fig. \ref{fig:flow}.
	One of the advantages of the proposed quasi-1D model is that no iteration or discretization is involved in the calculation, enabling a very rapid computation.
	For the test cases in this paper, the quasi-1D model is implemented in the MATLAB environment.
	The computation for one test case is completed within 1 s by using a quad-core desktop computer, using a resolution of 400 $\times$ 400 for the 2D distributions.

\section{Case Analysis I: Hydrogen-Air RDC}
	\label{sec:case1}
	
\subsection{Distribution Comparison}
\begin{figure*}[b]
		\begin{center}
			\includegraphics[width=160mm]{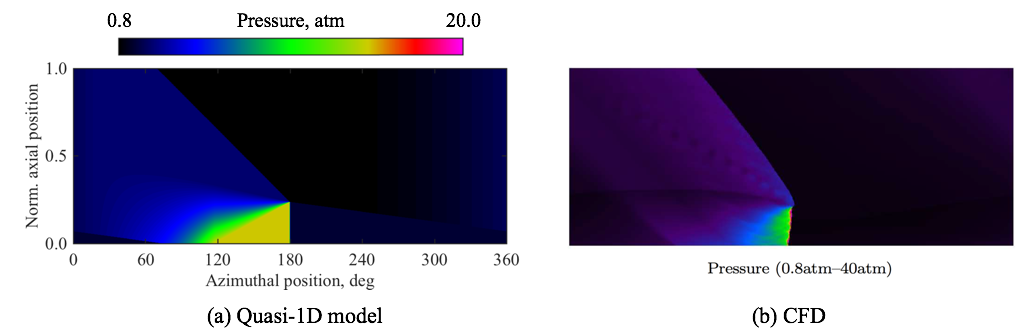}
		\end{center}
		\caption{Annulus pressure distribution simulated with the (a) quasi-1D model and (b) CFD [\citen{Schwer:2010aa}] for the hydrogen-air case.
		}
		\label{fig:pres1}
	\end{figure*}

\begin{figure*}[t]
		\begin{center}
			\includegraphics[width=160mm]{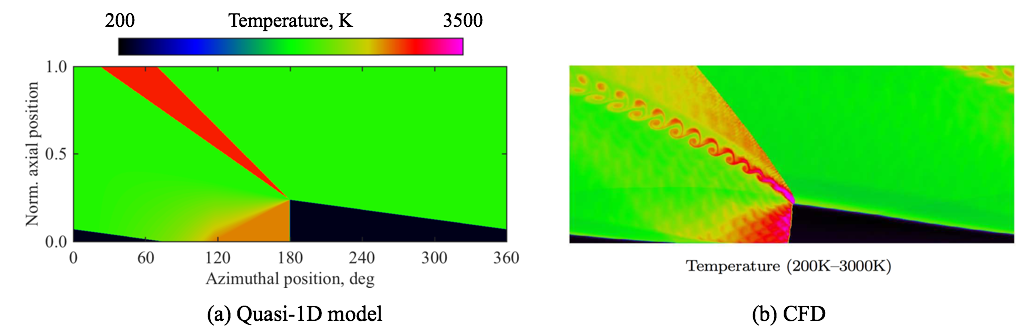}
		\end{center}
		\caption{Annulus temperature distribution simulated with the (a) quasi-1D model and (b) CFD [\citen{Schwer:2010aa}] for the hydrogen-air case.
		}
		\label{fig:temp1}
	\end{figure*}

	A case analysis has been conducted for the RDC using hydrogen as the fuel and air as the oxidizer.
	The input parameters used in this analysis are listed in Table 1.
	These input parameters are consistent with the calculation condition of the CFD analysis conducted by Schwer and Kailasanath [\citen{Schwer:2011aa,Schwer:2010aa}].
	The calculated two-dimensional distributions of pressure and temperature are shown in Figs. \ref{fig:pres1} and \ref{fig:temp1}, respectively.
	In a qualitative sense, the pressure and temperature distributions calculated by the quasi-1D model are consistent with those obtained by CFD.
	The wave structure of the shock wave, contact surface, and expansion wave calculated in the quasi-1D model is similar to that simulated in CFD.
	
\begin{figure*}[t]
		\begin{minipage}{0.5\hsize}
		\begin{center}
			\includegraphics[width=80mm]{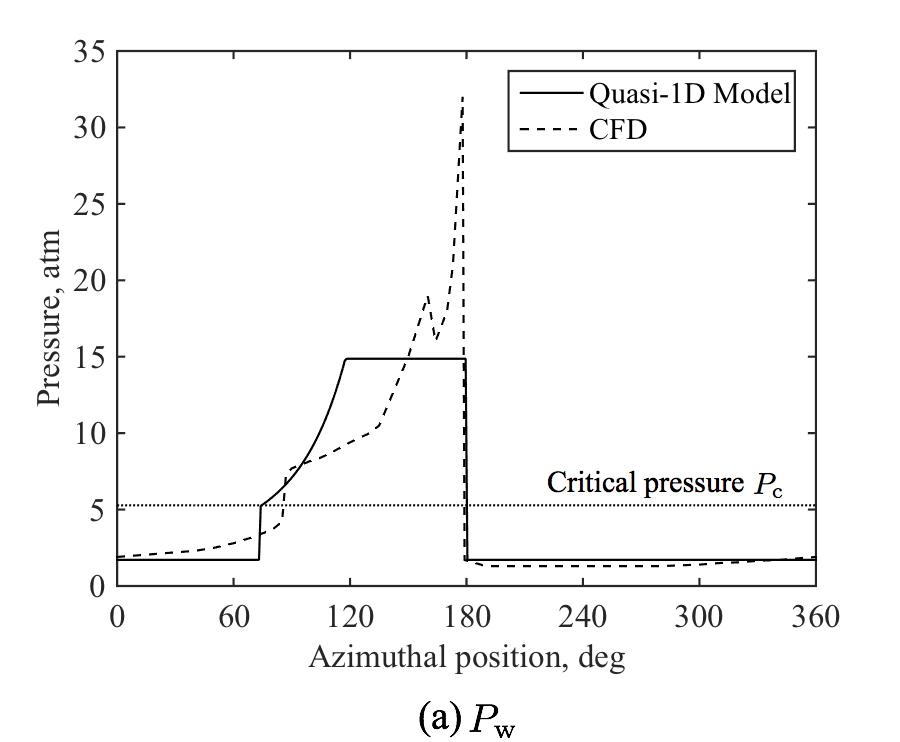}
		\end{center}
		\end{minipage}
		\begin{minipage}{0.5\hsize}
		\begin{center}
			\includegraphics[width=80mm]{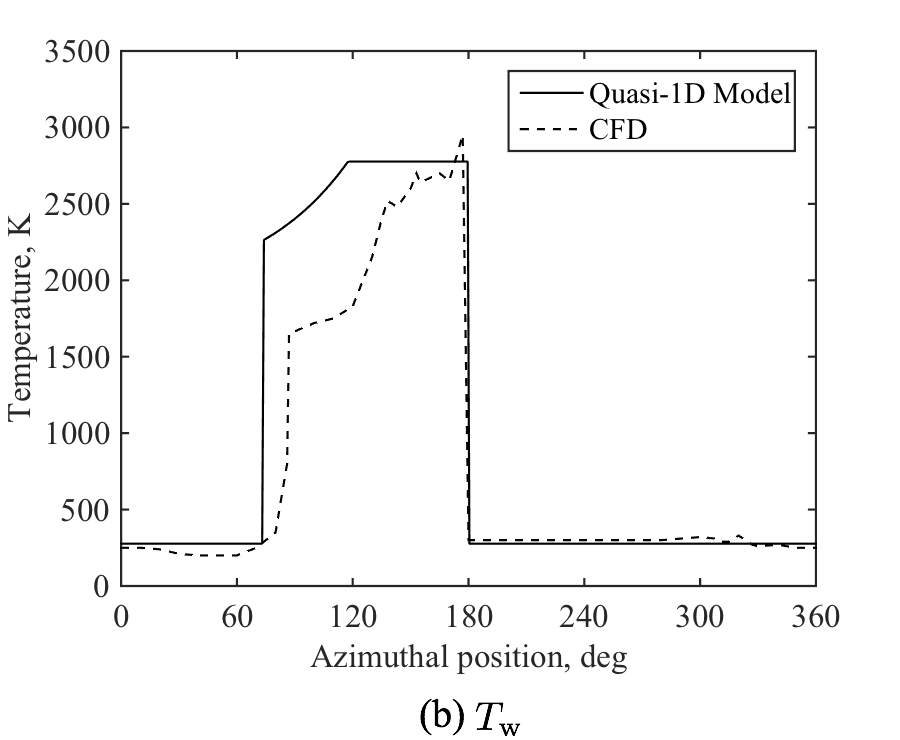}
		\end{center}
		\end{minipage}
		\caption{Azimuthal distributions of (a) $P_{\rm w}$ and (b) $T_{\rm w}$ calculated by the quasi-1D model and CFD [\citen{Schwer:2010aa}].
		}
		\label{fig:wall1}
	\end{figure*}

	The azimuthal distributions of the wall pressure $P_{\rm w}$ calculated by the quasi-1D model and CFD are compared in Fig. \ref{fig:wall1}(a).
	The result of CFD exhibits a sharp peak, the von Neumann spike, which is not captured in the result of the quasi-1D model.
	This is because the quasi-1D model confines the azimuthal flow, and the flow field induced behind the detonation wave cannot be taken into account by the model.
	Instead of the von Neumann spike, the pressure distribution of the quasi-1D model is characterized by the plateau region.
	The plateau region stems from the assumption of the shock-tube model.
	After the plateau region, the pressure rapidly decreases in accordance with the pressure decay function.
	Once the pressure reaches the critical pressure $P_{\rm c}$, the propellant injection is initiated, and the pressure drops to $P_0$.
	In addition, the temperature distribution along the wall is shown in Fig. \ref{fig:wall1}(b).
	In the results of both quasi-1D model and CFD, the temperature behind the detonation is approximately 2700 K.
	The pressure and temperature ahead of the detonation wave show agreement between the results of the quasi-1D model and CFD.
	
\subsection{Performance Comparison}
	For a quantitative comparison, the thruster performances calculated by the quasi-1D model and CFD are compared.
	The dependence of the thruster performance on the plenum pressure $P_{\rm p}$ is investigated by calculating the five cases of $P_{\rm p}=$5, 7.5, 10, 15, and 20 atm.
	
\begin{figure*}[t]
		\begin{center}
			\includegraphics[width=160mm]{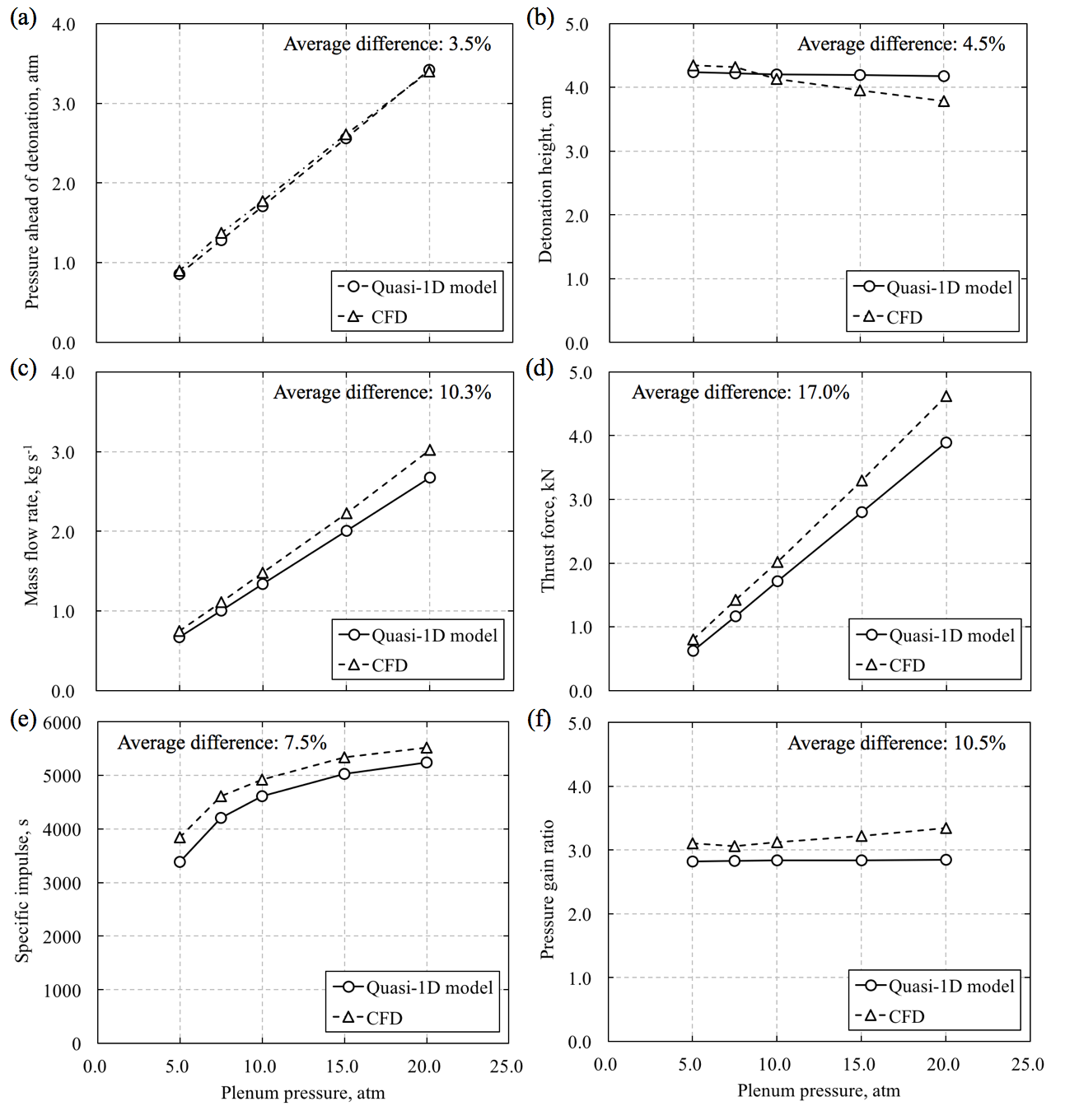}
		\end{center}
		\vspace{-5mm}
		\caption{Thruster performances calculated by the quasi-1D model and CFD [\citen{Schwer:2010aa}].
		(a) $P_0$, (b) $h_{\rm det}$, (c) $\dot{m}$, (d) $F$, (e) $I_{\rm sp}$, and (f) $r_{\rm PG}$.
		}
		\label{fig:performance}
	\end{figure*}

	Fig. \ref{fig:performance} compares various quantities indicating the thruster performances calculated by the quasi-1D model and CFD.
	The pressure ahead of detonation $P_0$ derived in the model is quite close to that calculated in CFD, as shown in Fig. \ref{fig:performance}(a).
	Both the quasi-1D model and CFD showed a weak dependence of the detonation height $h_{\rm det}$ on the plenum pressure, as shown in Fig. \ref{fig:performance}(b).
	The difference of $h_{\rm det}$ between the quasi-1D model and CFD is within 5\%, which indicates that the detonation wave height model used in Sec. \ref{sec:wall} is reasonable.
	
	As shown in Figs. \ref{fig:performance}(c) and (d), the mass flow rate and thrust force calculated in the quasi-1D model are slightly underestimated compared with the results of CFD.
	Nevertheless, the differences of $\dot{m}$ and $F$ between the results of the quasi-1D model and CFD are within 20\%, and the linear trends are well reproduced in the quasi-1D model.
	Fig. \ref{fig:performance}(e) compares the specific impulse $I_{\rm sp}$.
	$I_{\rm sp}$ derived in the quasi-1D model shows a good agreement with the result of CFD, with an average difference of 7.5\%.
	Further, the curved trend of $I_{\rm sp}$ dependence on the plenum pressure is well reproduced, which supports the validity of the quasi-1D model.
	
	Both the quasi-1D model and CFD indicate a pressure gain ratio $r_{\rm PG}$ of approximately 3, as shown in Fig. \ref{fig:performance}(f).
	This result proves the occurrence of pressure gain combustion in the RDC of the simulated case.
	It is also proved that $r_{\rm PG}$ has little dependence on the plenum pressure.

	\begin{table}[t]
		\begin{center}
			\caption{Input parameters for the hydrogen-air RDC analysis.}
			{\begin{tabular}{p{45mm}p{15mm}p{15mm}}
				\Hline
				 Parameter & Symbol & Value \\
				 \hline
				 Fuel     & -- & Hydrogen \\
				 Oxidizer & -- & Air \\
			    Azimuthal annulus length, mm & $L_{\theta}$ & 439.8  \\
			    Axial annulus length, mm & $L_{\rm c}$ & 177.0  \\
				 Annulus width, mm & $w_{\rm c}$ & 20.0\\
			    Plenum pressure, atm & $P_{\rm p}$ & 10.0  \\
				 Plenum temperature, K & $T_{\rm p}$ & 300 \\
				 Injector-wall area ratio & $A_{\rm i}/A_{\rm w}$ & 0.2\\
				 Equivalence ratio & -- & 1.0\\
 				 Atmospheric pressure, atm & $P_{\rm a}$ & 1.0\\
				 \Hline
			\end{tabular}}
		\end{center}
		\label{tab:case1}
	\end{table}

	
\section{Case Analysis II: Ethylene-Oxygen RDC}
	\label{sec:case2}

	Another case analysis has been conducted for the ethylene-oxygen RDC.
	The purpose of this analysis is to examine the range of applicability of the quasi-1D model.
	The input parameters used in this analysis are presented in Table 2.
	These input parameters are consistent with the calculation condition of the CFD analysis conducted by Fujii et al. [\citen{Fujii:2017aa}].
	The quasi-1D model used in this analysis is identical to the one used in Case Analysis I.
	It is noted that the distributions within the axial position of 0--22.8 mm are used for the comparison, although the axial annulus length is 123 mm.
	This is because the grid sizing becomes coarse in the CFD analysis in the downstream region beyond the axial position of 22.8 mm.
	
\begin{figure*}[h]
		\begin{center}
			\includegraphics[width=160mm]{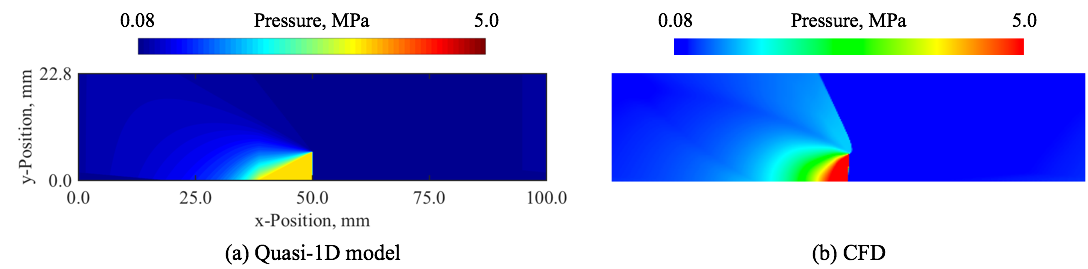}
		\end{center}
		\vspace{-4mm}
		\caption{Annulus pressure distribution simulated with the (a) quasi-1D model and (b) CFD [\citen{Fujii:2017aa}] for the ethylene-oxygen case.
		}
		\label{fig:pres2}
		\vspace{50mm}
		\begin{center}
			\includegraphics[width=160mm]{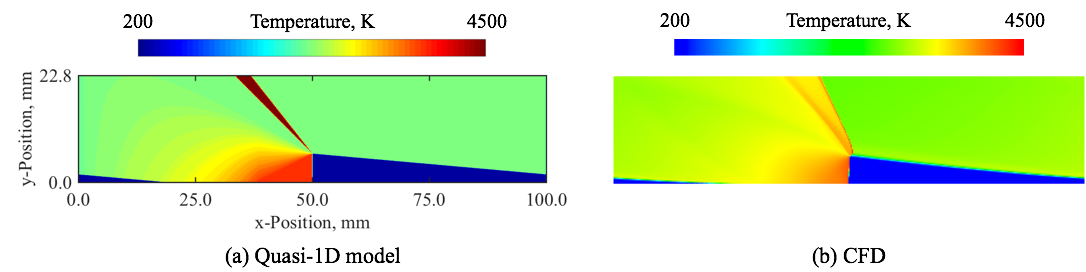}
		\end{center}
		\vspace{-4mm}
		\caption{Annulus temperature distribution simulated with the (a) quasi-1D model and (b) CFD [\citen{Fujii:2017aa}] for the ethylene-oxygen case. 
		}
		\label{fig:temp2}
	\end{figure*}

	The calculated two-dimensional distributions of pressure and temperature are shown in Figs. \ref{fig:pres2} and \ref{fig:temp2}, respectively.
	The two-dimensional distributions obtained by the quasi-1D model are qualitatively consistent with those of the CFD analysis.
	The detonation wave height $h_{\rm det}$ calculated by the quasi-1D model is 6.2 mm, whereas $h_{\rm det}$ is estimated as 6.1 mm from the image analysis in the CFD result.
	Thus, the quasi-1D model accurately reproduces $h_{\rm det}$, the difference being within 2\% compared to the CFD.
	It is proved that the quasi-1D model also yields distributions that are consistent with the high-fidelity CFD in the case of the ethylene-oxygen RDC, in addition to the hydrogen-air RDC.
	This fact expands the range of applicability of the quasi-1D model for various types of propellants and combustor geometries.

	\begin{table}[t]
		\begin{center}
			\caption{Input parameters for the ethylene-oxygen RDC analysis.}
			{\begin{tabular}{p{45mm}p{15mm}p{15mm}}
				\Hline
				 Parameter & Symbol & Value \\
				 \hline
				 Fuel     & -- & Ethylene \\
				 Oxidizer & -- & Oxygen \\
			    Azimuthal annulus length, mm & $L_{\theta}$ & 100.0  \\
			    Axial annulus length, mm & $L_{\rm c}$ & 123.0  \\
				 Annulus width, mm & $w_{\rm c}$ & 20.0\\
			    Plenum pressure, atm & $P_{\rm p}$ & 10.0  \\
				 Plenum temperature, K & $T_{\rm p}$ & 293\\
				 Injector-wall area ratio & $A_{\rm i}/A_{\rm w}$ & 0.2\\
				 Equivalence ratio & -- & 1.0\\
 				 Atmospheric pressure, atm & $P_{\rm a}$ & 1.0\\
				 \Hline
			\end{tabular}}
		\end{center}
		\label{tab:case2}
	\end{table}

\section{Conclusion}
	A quasi-1D model is proposed for the internal fluid of annulus RDCs.
	This model applies the micro-shock-tube model that constrains the flow only in the axial direction, while admitting detonation wave propagation in the azimuthal direction.
	The wall pressure distribution is considered on the basis of the analytic model of the PDE, and a theoretical estimation method for the detonation height is proposed.
	The proposed model can quickly analyze the thruster performance and 2D distributions of the gas properties because no iteration or discretization is involved.
	
	The validity of the model is examined by comparing the results calculated by the quasi-1D model and CFD.
	Two case analyses of hydrogen-air and ethylene-oxygen RDCs are conducted.
	The distributions of pressure and temperature are qualitatively compared and the thruster performances are quantitatively compared.
	The findings are summarized as follows.
	1) The 2D pressure and temperature distributions calculated by the model are consistent with those simulated by CFD.
	The wave structure observed in CFD is maintained in the quasi-1D model.
	2) Good agreement has been confirmed between the thrust performances obtained by the quasi-1D model and CFD.
	For instance, the difference between the specific impulses calculated by the model and CFD is 7.5\%.
	3) Consistency of 2D distributions between the quasi-1D model and CFD has been observed in the ethylene-oxygen RDC case, in addition to the hydrogen-air RDC case. 
	This fact expands the range of applicability of the quasi-1D model for various types of propellants.

\section*{Acknowledgment}
	This paper is based on results obtained from a project commissioned by the New Energy and Industrial Technology Development Organization (NEDO).
	The authors would like to thank Dr. Yusuke Maru (Japan Aerospace Exploration Agency) and Dr. Ken Matsuoka (Nagoya University) for valuable comments.

\end{document}